\documentclass{Interspeech}

\interspeechcameraready

\title{Speechless: Speech Instruction Training Without Speech for Low Resource Languages}

\author[affiliation={1}]{Alan Dao (Gia Tuan Dao)}{}
\author[affiliation={1}]{Dinh Bach}{Vu}
\author[affiliation={1}]{Huy Hoang}{Ha}
\author[affiliation={1}]{Tuan}{Le Duc Anh}
\author[affiliation={2}]{Shreyas}{Gopal}
\author[affiliation={2}]{Yue Heng}{Yeo}
\author[affiliation={1}]{Warren Keng Hoong}{Low}
\author[affiliation={2}]{Eng Siong}{Chng}
\author[affiliation={1}]{Jia Qi}{Yip}

\affiliation{}{}{Menlo Research}
\affiliation{CCDS}{Nanyang Technological University}{Singapore}
\email{alan@menlo.ai}


\usepackage{comment}

\begin{document}

\maketitle

\begin{abstract}
     The rapid growth of voice assistants powered by large language models (LLM) has highlighted a need for speech instruction data to train these systems. Despite the abundance of speech recognition data, there is a notable scarcity of speech instruction data, which is essential for fine-tuning models to understand and execute spoken commands. Generating high-quality synthetic speech requires a good text-to-speech (TTS) model, which may not be available to low resource languages. Our novel approach addresses this challenge by halting synthesis at the semantic representation level, bypassing the need for TTS. We achieve this by aligning synthetic semantic representations with the pre-trained Whisper encoder, enabling an LLM to be fine-tuned on text instructions while maintaining the ability to understand spoken instructions during inference. This simplified training process is a promising approach to building voice assistant for low-resource languages. 
\end{abstract}

\section{Introduction}
Voice assistants have become an integral part of modern technology, providing users with the ability to interact with devices through natural language~\cite{dao2024ichigo}. These voice assistants can be achieved through a cascade of automatic speech recognition (ASR) which transcribes instructions which are then processed by an instruction-tuned large language model (LLM)~\cite{ji2024wavchat}. However, in such a cascaded implementation, the latency introduced by the ASR model can negatively impact the user experience. Thus, early-fusion models~\cite{cui2024recent}, where the language model is fine-tuned to accept speech representations instead of ASR transcripts, have become an increasingly popular solution. However, this fine-tuning process requires a significant amount of speech instruction and its corresponding response data. For example, LLaMA-Omni~\cite{fang2024llama} was trained on the InstructS2S-200k~\cite{fang2024llama} dataset, which consists of 200k speech instructions and their corresponding speech responses. These speech instructions are linguistically different from ASR data as they consists of questions and answers, whereas ASR transcripts consists primarily of statements. As such, these types of spoken instructions are scarce compared to ASR data even for common languages, and the problem is more acute for low resource languages like Vietnamese.

The most cost-effective method for tackling the lack of spoken instruction data is to generate synthetic data. Researchers most commonly make use of text to speech (TTS) systems to generate speech~\cite{noroozi24InstructionDataGeneration}. Given a dataset of questions and answers, the questions in the dataset can be sent to a TTS model to generate spoken questions~\cite{majumder2024tango}. In some cases, the text of the questions are also generated by a prior LLMs~\cite{tang2024salmonn,pan2023cosmic,zhao2023librisqa}. While this approach has been shown by~\cite{noroozi24InstructionDataGeneration} to be broadly effective, the method is reliant on the quality of the TTS models used, and requires a diversity of speaker voices to be simulated for good generalization. While high resource languages like English can benefit from high quality TTS models, TTS models of low resource languages such as Vietnamese can often lag behind in performance~\cite{tran22_interspeech, lux-etal-2022-low}. Without TTS, some methods~\cite{manakul2024enhancing, chung2018unsupervised} have relied on text-audio alignment using ASR data to avoid the need for spoken instruction fine-tuning~\cite{held2024distilling, huzaifah2023analysis, gaur24_interspeech}. 

\begin{figure}[t]
    \centering
    \includegraphics[width=0.8\linewidth]{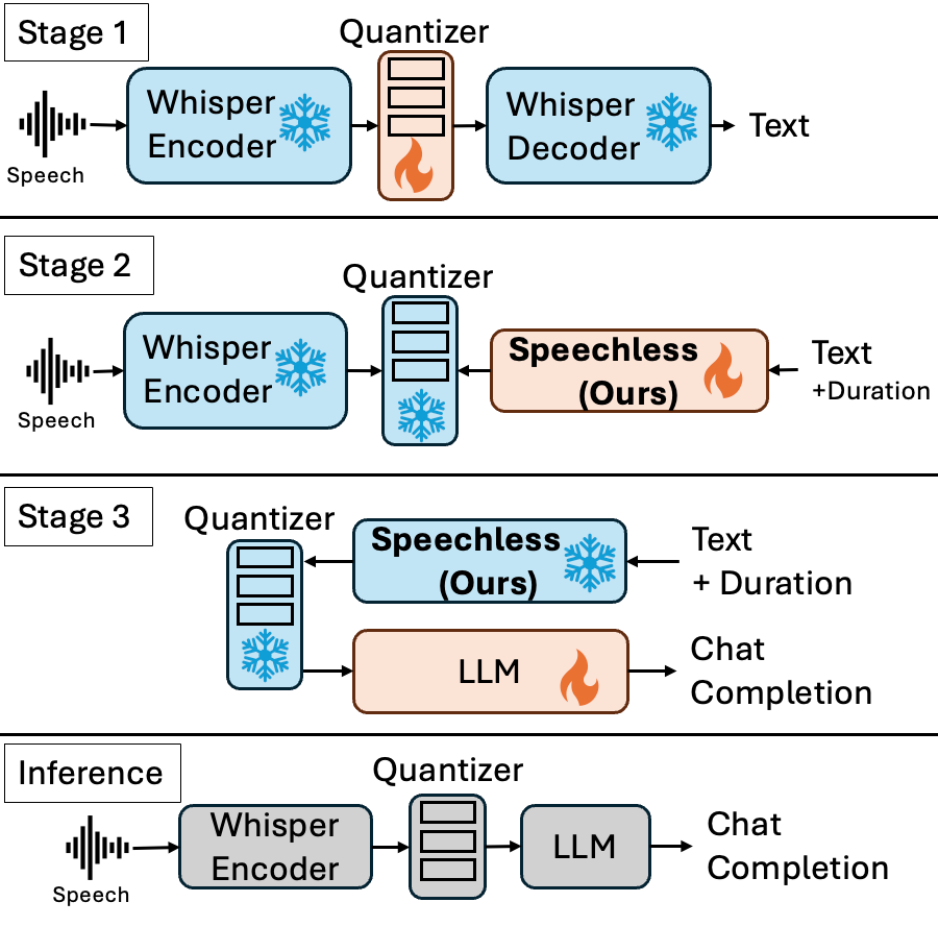}
    \caption{Overview of the training process using Speechless. In Stage 1, we train a quantizer using ASR data. In Stage 2, we train Speechless, which maps text and duration tokens to audio tokens. In Stage 3, we fine-tune an LLM using audio tokens generated by Speechless. At inference time the LLM is able to accept speech input through the Whisper Encoder, even though no speech data was used to fine-tune the LLM.}\label{fig:overview}
    \vspace{-15 pt}
\end{figure}

In this work, we propose \textit{Speechless}, a novel method for generating synthetic training data for early-fusion speech language models without relying on traditional TTS systems. As illustrated in Figure~\ref{fig:overview}, Speechless generates semantic speech tokens using a quantized Whisper encoder~\cite{radford2023robust}, bypassing the need for waveform generation entirely. By leveraging the Whisper encoder's inherent noise robustness and speaker invariance at inference time, our approach avoids the need for speaker diversity. Meanwhile, semantic diversity can be efficiently achieved using LLMs and readily available text corpora. Furthermore, since the speech encoder remains frozen during training, fine-tuning is performed exclusively at the token level, significantly reducing computational costs compared to traditional speech-based fine-tuning pipelines.

Our contributions are threefold: First, we propose \textit{Speechless}, a novel paradigm that generates instruction training data through semantic token alignment with Whisper's encoder, eliminating dependency on TTS systems. Second, we demonstrate this method's effectiveness for Vietnamese - a language with limited TTS resources - achieving competitive ASR performance without speech-based fine-tuning. Third, we release the first pre-tokenized Vietnamese instruction dataset enabling speech-language model development, addressing a critical gap in low-resource NLP. 

\section{Methodology}

Our method consists of three main stages, as illustrated in Figure 1. First, we train a residual vector quantizer (RVQ) to encode speech into discrete semantic tokens that align with Whisper's encoder representations. Second, we develop Speechless, a decoder-only language model that learns to generate these semantic tokens directly from text, effectively creating a text-to-semantics model that bypasses the need for audio generation. Finally, we use Speechless to generate synthetic training data for fine-tuning LLMs on speech understanding tasks. 

This approach allows us to create high-quality training data without relying on traditional text-to-speech systems, making it particularly valuable for low-resource languages where such systems may be limited or unavailable. Our method builds on earlier works~\cite{Ao2021SpeechT5, huzaifah2023analysis, gaur24_interspeech, held2024distilling} that have aimed to align speech and text modalities, but is aimed towards low resource languages and tries to leverage the large-scale pretraining of Whisper~\cite{radford2023robust}. The training code for all three stages are released on GitHub~\footnote{\href{https://github.com/menloresearch/ichigo/tree/legacy/main}{https://github.com/menloresearch/ichigo/tree/legacy/main}}.

\subsection{Stage 1: Training a Quantizer}

Stage 1 of Speechless focuses on training a quantizer that aligns the semantic and text representations, making the downstream task of training Speechless easier. At the core of this stage is a residual vector quantizer (RVQ), which transforms the high-dimensional speech representations from Whisper's encoder into discrete tokens while preserving semantic meaning. RVQ achieves this through an iterative refinement process: First, it creates a coarse representation of the input using an initial codebook, then progressively refines this representation by quantizing the residual errors through subsequent codebooks. This multi-stage approach allows the quantizer to capture both broad semantic features and subtle nuances in speech.

To adapt the quantizer for low-resource languages, we expanded the quantizer's capacity by quadrupling the codebook size from 512 to 2048 entries. Our initial attempt to initialize the expanded codebook using Kaiming initialization led to poor codebook utilization. To overcome this, we adopted a different strategy by duplicating the original codebook weights and applying Kaiming-initialized random noise to these duplicates.

\subsection{Stage 2: Training Speechless}
Speechless is a 1 billion parameter decoder-only language model designed to generate semantic representations of audio as discrete tokens. By treating semantic tokens as a novel language, Speechless functions similarly to a machine translation model. It translates text-based instructions into a sequence of semantic tokens that is close to what would be generated by the Whisper Encoder if the same text had been spoken, recorded and played into the Whisper Encoder. This close alignment between the Speechless output and Whisper Encoder output allows us to train using only text instructions, but have the model understand speech at inference.

A key challenge in text-to-speech conversion is managing the mismatch between text and speech tokens, since a given text sequence typically corresponds to significantly more speech tokens, and the number of speech tokens needed can vary. Speechless addresses this challenge through its design as an auto-regressive decoder model that predicts tokens one at a time. This auto-regressive approach allows the model to flexibly generate the appropriate number of speech tokens for any input text, regardless of length. Additionally, we provisioned the model with a billion parameters to give it a sufficiently large vocabulary for this task. While this is a large number of parameters for a speech model, this model is not used during inference and thus does not impact inference cost.

Speechless accepts text instructions from standard LLM instruction datasets and outputs semantic tokens through the Ichigo tokenizer, building upon Ichigo's \cite{dao2024ichigo} successful use of semantic tokens for instruction-response pairs. To train Speechless, instead of QA pairs, we used transcription text and sematic tokens pairs. This approach mirrors the dynamics of machine translation models, where the model learns to map structured inputs (text instructions) to meaningful outputs (semantic tokens) through extensive training on paired data. The semantic tokenization process abstracts away the acoustic details, focusing instead on the underlying meaning, which allows for robust and flexible applications across diverse languages and contexts.

To train Speechless, we utilize speech-text pairs from an ASR dataset. The raw transcripts serve as input to Speechless, appended after a special task token $<|text\_to\_semantic|>$. Speech is tokenized using the quantizer from Stage 1 and serves as the target output. To reduce computational costs, we compress the target sequence length by creating a $<|duration|>$ token to represent the repetition inside each repeated sound tokens group. 

\subsection{Stage 3: Training the LLM}
After training Speechless, we use it to generate synthetic data that can be used to fine-tune a pre-trained LLM. For text data, we combined multiple instruction datasets: Ichigo ~\cite{dao2024ichigo} for English content, and Sailor ~\cite{Sailor2} and Viettel x NVIDIA ~\cite{ViettelXNvidia} datasets for Vietnamese. The data preparation process involved several filtering steps to ensure quality: we removed samples with excessive prompt lengths, filtered out non-audible content (such as mathematical equations and excessive punctuation), and curated responses by refining the answers for the Viettel dataset using the Qwen2.5-32B model. Finally, we tokenize the user turn to discrete speech tokens using Speechless.

Subsequently, after the synthetic semantic tokens have been generated, we can apply a standard speech instruction tuning pipeline~\cite{taori2023alpaca} with minimal modifications. To ensure that training with Speechless was successful, we added new sound and duration tokens to the LLaMA tokenizer and resized the embedding and finally linear head of the LLaMA model, so that the model could train with the new tokens. Thus, by adding Speechless into any instruction tuning pipeline, we can train the model with only text instructions, but have the model understand speech at inference.

\begin{table*}[h]
    \captionsetup{skip=0pt}
    \centering
    \caption{All results are in percentages. Comparative analysis of model performance for general, noisy, and multilingual ASR using the LibrisSpeech (LS), VoiceBank+DEMAND (VBD), and CommonVoice (CV) datasets respectively. All results are derived from processed labels and predictions. Both labels and predictions are lower-cased and all special characters are removed.}
    \begin{tabular}{llcccccccccc}
        \toprule
        Model & Config & \multicolumn{2}{c}{LS test-clean} & \multicolumn{2}{c}{VBD clean} & \multicolumn{2}{c}{VBD noisy} & \multicolumn{2}{c}{CV En} & \multicolumn{2}{c}{CV Vi} \\
        \cmidrule(lr){3-4} \cmidrule(lr){5-6} \cmidrule(lr){7-8} \cmidrule(lr){9-10} \cmidrule(lr){11-12}
        & & CER & WER & CER & WER & CER & WER & CER & WER & CER & WER \\
        \midrule
        \textbf{Whisper (M)} & Zero-shot (greedy) & 1.21 & 2.85 & 1.45 & 4.99 & 2.13 & 6.17 & 4.21 & 5.98 & 15.00 & 25.43 \\
        \textbf{} & Zero-shot (beam-10) & 0.92 & 2.51 & 1.33 & 4.80 & 1.94 & 5.91 & 3.21 & 5.22 & 13.72 & 24.18\\
        \midrule
        \textbf{Whisper (M)} & Greedy Inference & 3.45 & 6.74 & 3.27 & 7.12 & 9.32 & 15.76 & 4.33 & 7.27 & 28.11 & 36.53 \\
        \textbf{Quantized}& Beam-Search (n=10) & 2.42 & 5.52 & 2.89 & 6.76 & 6.63 & 12.34 & 3.24 & 7.01 & 24.16 & 34.84 \\
        \midrule
        \textbf{Speechless} & Greedy Inference & 2.47 & 4.65 & 1.01 & 2.32 & - & - & 3.54 & 8.03 & 2.69 & 5.90 \\
         & Beam-Search (n=10) & 2.08 & 4.21 & 1.52 & 3.92 & - & - & 2.92  & 6.56 & 3.77 & 7.08 \\
        \bottomrule
    \end{tabular}
    \label{tab:model_results}
\end{table*}

\section{Experiments}
\subsection{Datasets}
For Stage 1, we utilized two automatic speech recognition (ASR) datasets: viVoice (Vietnamese) and LibriTTS-R\cite{zen2019libritts} (English). The ViVoice dataset consists of 868k utterances for training, 10k for validation, and 10k for testing, while the LibriTTS-R dataset contains 112k training samples, 5.6k validation samples, and 4.6k test samples. Since the training data primarily consisted of clean speech, the resulting model exhibited increased sensitivity to noise.

For training Stage 2, we took 880k samples from Vivoice~\cite{viVoice}, 112k samples from LibriTTS-R Clean~\cite{kawamura2024librittspcorpusspeakingstyle} and converted the audios from these datasets into semantic tokens using the quantizer in Stage 1. Then, we use transcriptions of the corresponding quantized audio to create text-to-semantic training pairs.

For Stage 3, we took 880k samples from Vivoice~\cite{viVoice}, 112k samples from LibriTTS-R Clean~\cite{kawamura2024librittspcorpusspeakingstyle}, and 2.4M samples from MLS Eng~\cite{Pratap2020MLSAL} 10k as our pretraining data, and converted the transcripts from these datasets into semantic tokens using the quantizer and the Speechless model. We then used Ichigo's~\cite{dao2024ichigo} instruction data to train the model on instructions, using Speechless~\footnote{\href{https://huggingface.co/Menlo/Speechless-llama3.2-v0.1}{https://huggingface.co/Menlo/Speechless-llama3.2-v0.1}} to convert the audio into semantic tokens. We release this synthetic dataset on Huggingface~\footnote{\href{https://huggingface.co/datasets/Menlo/Ichigo-instruction-tokenized-v0.2}{https://huggingface.co/datasets/Menlo/Ichigo-instruction-tokenized-v0.2}}.

To create the synthetic data for Stage 3, we created an efficient pipeline using vLLM~\cite{vllm} to batch inference Speechless model which has only 1B (takes up to 4GB VRAM). We used Ray~\cite{moritz2018ray} for distributed processing, running vLLM~\cite{vllm} instances across multiple GPUs. 

\subsection{Base Models Used} 
For the Whisper models used in Stage 1 and 2 of the training, we start with a Whisper Medium checkpoint~\footnote{\href{https://huggingface.co/Menlo/Ichigo-whisper-v0.1}{https://huggingface.co/Menlo/Ichigo-whisper-v0.1}}. All input audio was padded to 30 seconds where applicable, to be compatible with the default implementation of Whisper. For Stage 3, we tested downstream finetuning with Speechless on both base and instruct versions of LLaMA 3.2 1B and LLaMA 3.2 3B. However, for most of our experimentation, we chose the LLaMA 3.2 1B Base model. Our preliminary experiments showed that the 3B model performed similarly to the 1B model, so we choose to use LLaMA 3.2 1B for resource efficiency.

\subsection{Training Cost}
The training process for our model was divided into three distinct stages, each with its own computational requirements. In Stage 1, the training was conducted over two phases. Phase 1 required 75 hours to complete 50 epochs, while Phase 2 took 29 hours for 20 epochs. This stage utilized 8 A6000 GPUs, with a batch size of 42 per GPU. The learning rate was set at 1e-3, using the AdamW optimizer. A linear warm-up was applied for the first 500 steps, followed by a cosine decay schedule, and a weight decay of 0.001 was implemented.

Stage 2 of the training process was completed in 60 hours, using 6 A6000 GPUs. The batch size was increased to 48 per GPU, and the learning rate was adjusted to 1e-4. Similar to Stage 1, a linear warm-up was used for the initial 100 steps, followed by a cosine decay schedule, with a weight decay of 0.01.

Finally, Stage 3 was divided into two parts: pretraining and supervised finetuning. The pretraining phase took 240 hours on A6000 GPUs, with a batch size of 42 per GPU and a learning rate of 2e-4. The same scheduling strategy as the previous stages was applied. The supervised finetuning phase required 40 hours on H100 GPUs, with a batch size of 32 per GPU and a learning rate of 3e-4. This stage also used a linear warm-up for 100 steps, followed by a cosine decay schedule, and a weight decay of 0.01. Overall, the training process was resource-intensive, reflecting the complexity and scale of the model development.

\begin{table*}[h]
\centering
\caption{VoiceBench~\cite{chen2024voicebench} Results. These are results based on spoken questions and text answers. Experiments other than ours were performed by the VoiceBench authors. SD-QA and CommonEval have a human audio source, while the rest use Google TTS.}
\begin{tabular}{l|c|c|c|c|c}
\hline
\textbf{Model Name} & \textbf{AlpacaEval} & \textbf{CommonEval} & \textbf{SD-QA} & \textbf{OpenBookQA} & \textbf{AdvBench} \\
\hline
Baichuan-Omni-1.5 & 4.50 & 4.05 & 43.40 & 74.51 & 97.31 \\
GLM-4-Voice & 3.97 & 3.42 & 36.98 & 53.41 & 88.08 \\
Qwen2-Audio & 3.74 & 3.43 & 35.71 & 49.45 & 96.73 \\
VITA-1.0 & 3.38 & 2.15 & 27.94 & 29.01 & 26.73 \\
Moshi & 2.01 & 1.60 & 15.64 & 25.93 & 44.23 \\
\hline
Whisper-v3-turbo+LLaMA-3.1-8B & 4.55 & 4.02 & 58.23 & 72.09 & 98.46 \\
LLaMA-Omni & 3.70 & 3.46 & 39.69 & 27.47 & 11.35 \\
Speechless-llama3.1-8B-instruct (Ours) & 3.86 & 2.51 & 35.00 & 26.15 & 62.88 \\
\hline
\end{tabular}
\label{tab:voicebench}
\end{table*}
\section{Results}
\subsection{ASR and Speechless Comparisons}

To evaluate the performance of the Speechless model alone, we make use of ASR test sets. To do this evaluation, we compare semantic tokens generated by Whisper Encoder from speech with the semantic tokens generated by Speechless from text. In both cases the semantic tokens are decoded by the same Whisper Decoder model. Ideally, the WER for Speechless and Whisper Encoder should be similar for clean datasets, and the WER for Speechless should be better for noisy dataset.

To evaluate the general ASR capability of the quantized Whisper model, we used the \textit{test-clean} split of the LibriSpeech dataset \cite{librispeech_2015} and the \textit{clean} test set of the VBDemand dataset \cite{voicebank_d_2016}. We also evaluated the models using the \textit{test-other} split of LibriSpeech and the \textit{noisy} test set of VBDemand. The \textit{test-clean} split comprises 2,620 utterances, totaling approximately 5.4 hours of clean read speech, while the \textit{test-other} split contains 2,939 utterances, corresponding to 5.1 hours of read speech. The VBDemand test set includes 824 utterances, with the noisy subset incorporating background noise from eight DEMAND \cite{demand_2013} noise classes at varying signal-to-noise ratios. The transcripts from these datasets were used to evaluate the semantic token quality produced by the Speechless model after de-quantization and decoding via the Whisper decoder.

For multilingual ASR evaluation, we utilized the Vietnamese (VI) and English (EN) subsets from Mozilla Common Voice 17 \cite{commonvoice}. Common Voice is a crowd-sourced dataset containing recordings with diverse accents, dialects, and recording conditions, making it well-suited for assessing multilingual performance. We selected the official test splits, which include 4,325 utterances in Vietnamese and 6,125 in English, amounting to approximately 8.3 hours and 12.6 hours of speech, respectively. The transcripts from both subsets were used to evaluate the semantic token quality generated by the Speechless model following de-quantization and decoding through the Whisper decoder. This evaluation framework enabled us to measure the model’s performance across different languages and linguistic complexities.

In our efforts to establish a shared semantic language between Whisper and the Speechless LM, we first show that the Speechless model is able to generate semantic tokens that, when decoded by the Whisper decoder, display a very low WER across multiple domains of English and Vietnamese text data in \ref{tab:model_results}. This shows that Speechless is able to map raw text information to clean speech in the latent space. This is also clear when see that with added noise (VBD noisy), the Whisper encoder starts to generate tokens that show poorer WER in comparison.

We can also observe that once quantized, the Whisper encoder's performance declines in both noisy and multilingual settings. We posit that this is primarily due to information being lost during the residual vector quantization operation. As the Whisper decoder module is not trained after quantization layers are added, it is not privy to the change in latent speech embeddings. Additionally, only clean English and Vietnamese speech are used in Stage-1 training when generating the codebook, hence the codes may not have been exposed to noisy training data.

\subsection{LLM benchmarking}
To evaluate the performance of our model utilizing Speechless synthetic data for speech instruction tuning, we utilize the VoiceBench~\cite{chen2024voicebench} subset of the AlpacaEval~\cite{alpacaeval}, CommonEval~\cite{commonvoice}, SD-QA~\cite{faisal-etal-2021-sd-qa}, OpenBookQA~\cite{openbookqa} and AdvBench~\cite{advbench}, where text-based QA pairs have been converted to spoken friendly instructions and read through a TTS model. As the results reported in Table~\ref{tab:voicebench} show, Speechless achieves comparable performance to Llama-Omni, which is also uses LLaMA-3.1-8B, but was trained on spoken voice instructions. However, the performance of Whisper-v3-turbo+LLaMA-3.1-8B~\cite{chen2024voicebench}, which is a cascaded model, is significantly better. This is likely due to the LLM of a cascaded model only having to understand a single modality of data, text, which allows it to fully utilize its pre-training. Our Speechless fine-tuned model also outperforms Moshi~\cite{défossez2024moshispeechtextfoundationmodel}, while also achieving comparable performance to VITA-1.0. Our model underperforms newer models such as Baichuan-Omni-1.5, GLM-4-Voice, Qwen2-Audio which use different text LLMs as a starting point.

\begin{table}[h]
\centering
\caption{MMLU and VMLU Benchmarks. These are text-based benchmarks for comparing the performance degradation due to speech instruction tuning}
\begin{tabular}{l|c|c}
\hline
\textbf{Model Name} & \textbf{MMLU} & \textbf{VMLU} \\
\hline
meta-llama3.1-8B-instruct & 69.40 & 50.69 \\
Speechless-llama3.1-8B-instruct & 62.27 & 43.22 \\
\hline
\end{tabular}
\label{tab:text-benchmarks}
\end{table}

Next, in Table~\ref{tab:text-benchmarks} we report the performance of the model on MMLU and VMLU, which are text question and answer benchmarks. We find that our instruction-tuned model exhibits some performance degradation compared to the base model, which is expected as the model now has to be able to accept both speech and text tokens using the same number of parameters. Similar performance degradation after speech instruction tuning have also been previously reported in~\cite{wang2024blsp} and attributed to catastrophic forgetting.

\section{Conclusion}
This paper introduced \textit{Speechless}, a novel method for generating synthetic training data for early-fusion speech language models without traditional text-to-speech systems. By leveraging a quantized Whisper encoder, Speechless generates semantic speech tokens, effectively addressing challenges in low-resource languages. Our experiments demonstrated competitive performance across various ASR settings and enabled effective speech instruction tuning of LLMs.

However, our approach has limitations. The performance degradation observed in text-based benchmarks suggests potential issues with catastrophic forgetting during speech instruction tuning. Additionally, while Speechless shows promise in clean and controlled environments, its robustness in highly noisy or diverse linguistic contexts requires further exploration. Nevertheless, \textit{Speechless} the methods described in this paper can in principle be applied to noisy data. Thus, our future work will focus on enhancing noise robustness and expanding applicability to a broader range of languages and dialects.


\bibliographystyle{IEEEtran}
\bibliography{mybib}

\end{document}